\documentclass[a4paper,oneside,final,notitlepage,onecolumn,12pt]{article}
\usepackage{amsfonts}
\usepackage{epsf}
\usepackage{graphicx}
\usepackage{amssymb,eso-pic}
\usepackage{latexsym}
\usepackage{tabularx}
\usepackage{amsxtra} 
\usepackage{hyperref}
\usepackage{t1enc}
\usepackage{amsmath,accents}

\usepackage{cancel}

\usepackage{ifpdf,epsfig,array,amsmath,amssymb}

\usepackage{psfrag} 
\psfrag{na}{\footnotesize $n^a$}   
\psfrag{ta}{\footnotesize $\sigma^a$}   
\psfrag{vna}{\footnotesize $\check{n}^a$}
\psfrag{vNa}{\footnotesize $\check{N}^a$}
\psfrag{scrW}{\footnotesize $\mycal{W}_{\rho_0}$}
\psfrag{t=t1}{\footnotesize $\Sigma_{\sigma_1}$}
\psfrag{t=t2}{\footnotesize $\Sigma_{\sigma_2}$}
\psfrag{r=all}{\footnotesize $\rho=const$}
\psfrag{hna}{\footnotesize $\hat{n}^a$}
\psfrag{tna}{\footnotesize $\tilde{n}^a$}
\psfrag{hab,Kab}{\footnotesize $h_{ab}, K_{ab}$}
\psfrag{thab,tKab}{\footnotesize $\tilde h_{ab}, \tilde K_{ab}$}
\psfrag{hhab,hKab}{\footnotesize $\hat h_{ab}, \hat K_{ab}$}
\psfrag{chab,cKab}{\footnotesize $\check{h}_{ab}, \check{K}_{ab}$}

\usepackage[normalem]{ulem}
\usepackage{color}
\definecolor{blue}{rgb}{0,0,1}
\definecolor{red}{rgb}{1,0,0}

\definecolor{DGREEN}{rgb}{0,0.7,0.3}
\definecolor{grey1}{rgb}{0.52, 0.52, 0.51}

\newcommand{\interior}[1]{\accentset{\smash{\raisebox{-0.12ex}{$\scriptstyle\circ$}}}{#1}\rule{0pt}{2.3ex}}
\newcommand{\instar}[1]{\accentset{\smash{\raisebox{-0.12ex}{$\scriptstyle\star$}}}{#1}\rule{0pt}{2.3ex}}


\DeclareFontFamily{OT1}{rsfs}{} \DeclareFontShape{OT1}{rsfs}{m}{n}{
<-7> rsfs5 <7-10> rsfs7 <10-> rsfs10}{}
\DeclareMathAlphabet{\mycal}{OT1}{rsfs}{m}{n}

\def\sc{{\hskip 3.5pt {{}^{{}^{{}_{{}_{\bowtie}}}}} \kern -8.pt{}}}  
\def\SC{{\hskip 3.5pt {{}^{{}^{{}^{{}_{{}_{\bowtie}}}}}} \kern -10.5pt{}}}

\DeclareMathAlphabet{\mathpzc}{OT1}{pzc}{m}{it}

\def\tf{\textit{\texttt{f}}}
\def\tg{\textit{\texttt{g}}}
\def\th{\textit{\texttt{h}}}

\def\d{{\rm d}}

\begin{document}

\newtheorem{theorem}{Theorem}[section]
\newtheorem{lemma}{Lemma}[section]
\newtheorem{proposition}{Proposition}[section]
\newtheorem{corollary}{Corollary}[section]
\newtheorem{conjecture}{Conjecture}[section]
\newtheorem{condition}{Condition}[section]
\newtheorem{example}{Example}[section]
\newtheorem{definition}{Definition}[section]
\newtheorem{remark}{Remark}[section]
\newtheorem{exercise}{Exercise}[section]
\newtheorem{axiom}{Axiom}[section]
\renewcommand{\theequation}{\thesection.\arabic{equation}} 

\author{{Istv\'an R\'acz}\,\thanks{ ~email: racz.istvan@wigner.mta.hu} 
\\ 
\\ 
{Wigner Research Center for Physics} \\  {H-1121 Budapest,  Hungary
}}

\title{Constraints as evolutionary systems}

\maketitle

\begin{abstract}
The constraint equations for smooth $[n+1]$-dimensional (with $n\geq 3$) Riemannian or Lorentzian spaces satisfying the Einstein field equations are considered. It is shown, regardless of the signature of the primary space, that the constraints can be put into the form of an evolutionary system comprised either by a first order symmetric hyperbolic system and a parabolic equation or, alternatively, by a  symmetrizable hyperbolic system and a subsidiary algebraic relation. In both cases the (local) existence and uniqueness of solutions are also discussed.

\end{abstract} 

\date


\section{Introduction}\label{introduction}
\setcounter{equation}{0}

In solving the Cauchy problem in Einstein's theory of gravity suitable initial data on a smooth $n$-dimensional manifold $\Sigma$, also refereed as an initial data surface, have to be chosen. The geometric part of this data can be represented by a pair of smooth fields $(h_{ij},K_{ij})$ on $\Sigma$ such that $h_{ij}$ is a Riemannian metric while $K_{ij}$ is a symmetric tensor field there. The constraint equations deduced from Einstein's equation $G_{ij}-\mycal{G}_{ij}=0$, with a source $\mycal{G}_{kl}$ of vanishing divergence, are (for more details see \cite{racz_tdfd})
\begin{align} 
-\epsilon\,{}^{{}^{(n)}}\hskip-1mm R + \left({K^{j}}_{j}\right)^2 - K_{ij} K^{ij} - 2\,\mathfrak{e}=0\,, \label{expl_eh}\\
D_j {K^{j}}_{i} - D_i {K^{j}}_{j} - \epsilon\,\mathfrak{p}_i=0\,,\label{expl_em}
\end{align}
where $\epsilon$ takes the value $+1$ for a Riemannian or $-1$ for a Lorentzian primary space, whereas ${}^{{}^{(n)}}\hskip-1mm R$ and $D_i$ denote the scalar curvature and the covariant derivative operator associated with $h_{ij}$, respectively. The source terms $\mathfrak{e}$ and $\mathfrak{p}_i$ in (\ref{expl_eh}) and (\ref{expl_em}) are given as
\begin{equation}\label{1+n_source}
\mathfrak{e}= n^k n^l\,\mycal{G}_{kl} \hskip4mm {\rm and} \hskip4mm \mathfrak{p}_{i}=\epsilon\,{h^{k}}_{i} n^l\, \mycal{G}_{kl} \,.
\end{equation}

\medskip

The above introduced universality might seem to be overwhelming especially as already in the Lorentzian case immediate conceptual issues emerge. These are rooted in the highly underdetermined character of the constraints. In particular, as it was emphasized in \cite{wald}, the constraint equations impose only $n+1$ relations on the $n(n+1)$ independent components of $h_{ij}$ and $K_{ij}$, and the main issue is to single out from among these $n(n+1)$ independent components (or from their functions) those $n^2-1$ ones which can be freely specified, and those $n+1$ ones which are determined by the constraints, and how to identify the ones which correspond to gauge transformations.

\medskip

A unique and almost exclusive approach applied as yet to solve the constraint equations is the conformal method \cite{lich,york0,york1, choquet,wald}. The first essential step in developing this method was made seven decades ago by Lichnerowicz \cite{lich}. He constructed initial data for the $n$-body problem in general relativity by applying a  conformally flat metric $h_{ij}$ and vanishing extrinsic curvature $K_{ij}$. While applying the method proposed by Lichnerowicz one replaces the fields $h_{ij}$ and the trace free part of $K_{ij}$ by $\varphi^{\tfrac{4}{n-2}}\,\tilde h_{ij}$ and $\varphi^{-2}\,\tilde K_{ij}$, respectively, where $\varphi$ is a smooth nowhere vanishing scalar field while $\tilde h_{ij}$ is an arbitrary Riemannian metric on $\Sigma$. Then, it was shown by Lichnerowicz that a semilinear elliptic equation can be deduced from the Hamiltonian constraint for the conformal factor $\varphi$ in the time symmetric case with vanishing 
extrinsic curvature. 
The second crucial step was made by York \cite{york0,york1} in the early seventies. He completed the method of Lichnerowicz by recasting the momentum constraint into a linear elliptic form, and putting thereby the full set of constraint equations into a coupled semilinear elliptic system. 

\medskip

It is important to emphasize that by making use of the general method developed by Lichnerowicz and York one could, in principle, determine all the possible initial data configurations in general relativity. Nevertheless, as the momentum and Hamiltonian constraints decouple if the trace of the extrinsic curvature ${K^{l}}_{l}$ is constant---as well as, the momentum $\mathfrak{p}_i$ is York scaled (as described, e.g.~in \cite{choquet})---most of the related developments are also based on this simplification. Therefore, almost all of the existence, non-existence, or uniqueness theorems concerning the underlying semilinear elliptic system do also make use of these simplifications. For an excellent comprehensive recent review on the Cauchy problem and on the conformal method, in particular, on those results applicable in the non-constant mean curvature case, see Ref.~\cite{choquet}. 

\medskip

The currently applied methods in determining initial data for binary black hole simulations are also based on the conformal approach and on the aforementioned simplifications. For instance, the most widely used method---based on the proposal of Bowen and York \cite{bowen_york}---assumes that the metric $h_{ij}$ is conformally flat and that $\Sigma$ is maximal. As these assumptions are strictly stronger than the ones guaranteeing that the Hamiltonian and momentum constraints decouple we should not take by surprise that it is not possible to generate all the physically adequate initial data by making use of the corresponding method. Indeed, it was pointed out by Garat and Price in \cite{garat_price} (see also \cite{kroon} for a more detailed argument), no conformally flat axisymmetric slicing of the Kerr black hole spacetime exists. Essentially this---along with the  lack of a direct control of the physical parameters of the initial data specifications as a consequence of the implicit character of the 
conformal method, and along with the fact 
that 
one has to have both inner and outer boundary conditions and at the inner one we have much less support based on intuition than at the outer one---may be responsible for the non-negligible spurious gravitational wave content of the spacetimes evolved from Bowen-York type initial data specifications (for a discussion on this issue see, e.g., section 3.4.4. of \cite{alcu}).

\medskip

All the aforementioned observations underline that any alternative proposal allowing a convenient involvement of generic, i.e.~non-conformally flat, metric $h_{ij}$ and generic type of extrinsic curvature $K_{ij}$ could be of obvious interest. In this paper these type of alternative methods will be proposed. In particular, it will be shown here that if the initial data surface $\Sigma$ is foliated by a smooth one-parameter family of homologous $n-1$-dimensional surfaces then regardless whether the primary space is Riemannian or Lorentzian the $1+n$ momentum constraint can be written as a first order symmetric hyperbolic system, whereas the Hamiltonian constraint may be viewed as a parabolic or an algebraic equation depending on our preferences on the separation of the controlled\,\footnote{As it was shown in \cite{racz_geom_cauchy} the Hamiltonian constraint may also be put into the form of an elliptic equation for a conformal factor that can be defined by introducing the conformal structure 
on the foliating codimension-one surfaces (see also \cite{racz_tdfd} for more details on the conformal structure). Note, however, that the elliptic--hyperbolic systems yielded for the combination of this elliptic equation for the conformal factor and the hyperbolic system deduced from the momentum constraint is ill-posed \cite{racz_geom_cauchy}. Therefore, this elliptic formulation will not be considered in this paper.} and freely specifiable variables in (\ref{expl_eh}) and (\ref{expl_em}). 

\medskip

A key observation here is that as the constraint equations are highly underdetermined their character depend on the choice we make for the variables to be solved for. As a trivial example consider the equation 
\begin{equation}
\partial_x^2 \tf + \partial_y^2 \tf + \partial_x \tg - \partial_y^2 \tg + \th = 0
\end{equation}
which is elliptic, parabolic or algebraic depending on whether it is intended to be solved for the function $\tf$, $\tg$ or $\th$, respectively.

\medskip

Remarkably in both of the aforementioned cases the coupled Hamiltonian and momentum constraints can be solved as evolutionary systems. As it will be shown in section \ref{constraints}, if the Hamiltonian constraint is solved for the lapse of the foliation of $\Sigma$, and the level surfaces of the foliation are of positive mean curvature, then  the Hamiltonian constraint can be seen to possess the form of a parabolic partial differential equation, and the coupled constraints form then a hyperbolic-parabolic system. Notably, the Hamiltonian constraint can also be solved algebraically for one of the components of $K_{ij}$. By eliminating this component, along with its derivatives tangent to the foliating surfaces, the momentum constraint, under suitable conditions, maintains its hyperbolic character. More precisely, it may be put into the form of a first order Friedrichs symmetrizable hyperbolic system. It is unique that in both cases a well-posed initial value problem can be seen to apply to the systems comprised either by a 
first order symmetric hyperbolic system and a parabolic equation or by a first order symmetrizable hyperbolic system and a subsidiary algebraic relation.

\section{The basic setup}\label{setup}
\setcounter{equation}{0}

As it has already been indicated some restrictions concerning the topology of $\Sigma$ have to be imposed. In particular, we shall assume the existence of a smooth Morse function $\rho: \Sigma\rightarrow \mathbb{R}$ that does not possess more than two isolated  non-degenerate critical points on $\Sigma$.\,\footnote{The critical points of a Morse function $\rho$ (at which $D_i\rho=0$) are know to be isolated and non-degenerate in the sense that the Hessian of $\rho$ is non-singular at those points. The index of a critical point is 
the number of the negative eigenvalues of the Hessian there \cite{milnor}.} This means that $\Sigma$ can be foliated by a one-parameter family of homologous codimension-one surfaces $\mycal{S}_\rho$---the latter denote the $\rho=const$ level surfaces---such that (apart from possible critical points) the gradient $D_i\rho$ does not vanish on $\Sigma$. 

Note that the above assumptions---in spite of the involved technicalities---permit the involvement of high variety of topological spaces. For instance, all the product spaces $\mathbb{R}\times \mycal{S}$ are allowed, where the factor $\mycal{S}$ is a codimension-one manifold with arbitrary topology. This product structure guarantee that the ``height function'' determined by the factor $\mathbb{R}$ on $\Sigma$ will be a Morse function with no critical point.

Notice that $\Sigma=\mathbb{R}^n$ belongs to the above set corresponding to the particular choice $\mycal{S}=\mathbb{R}^{n-1}$. Note, however, that $\mathbb{R}^n$ may also be foliated---except at the origin---by codimension-one spheres $\mathbb{S}^{n-1}$ that corresponds to the Morse function $\rho=\sum_{i=1}^n (x_i)^2$ with zero index at the origin in $\mathbb{R}^n$. Note also that the above assumptions permit the involvement of the $n$-dimensional sphere $\mathbb{S}^n$ foliated by codimension-one spheres $\mathbb{S}^{n-1}$. This foliation corresponds to the choice of the height function $\rho(x_1,\dots,x_{n+1}) \mapsto x_{n+1}$, where $\mathbb{S}^n=\{(x_1,\dots,x_{n+1})\in\mathbb{R}^{n+1}\,|\, \sum_{i=1}^{n+1} (x_i)^2=\mathcal{R}^2 \}$ and the critical points are the north and south poles are represented by the points $(0,\dots,0,\mathcal{R})$ and $(0,\dots,0,-\mathcal{R})$ in $\mathbb{R}^{n+1}$, respectively.

In each of the above discussed cases the $\rho=const$ level surfaces are, or they are supposed to be, orientable either compact and without boundary in $\Sigma$ or non-compact and infinite.\,\footnote{The $\mycal{S}_\rho$ level surfaces could also be allowed to be compact with boundary in $\Sigma$. Note that in this case---which, for conveniences, is left out of the scopes of the 
present 
paper---both the hyperbolic-parabolic and the  symmetrizable hyperbolic systems can be 
solved (locally) as an initial boundary value problem such that the boundary values have to be fixed at the boundaries of the $\mycal{S}_\rho$ level surfaces  in $\Sigma$. 
(For useful hints on the solvability of the parabolic problem in the pertinent case see, e.g.~pages 25-42 in \cite{lsu}.)\label{FN}} 

\medskip

The rest of this section is to decompose---based on the use of the foliation $\mycal{S}_\rho$ of $\Sigma$---the Riemannian metric  $h_{ij}$ and the primary extrinsic curvature $K_{ij}$. 

\medskip

For instance, by making use of the unit norm vector field $\hat n^i$ normal to the $\mycal{S}_\rho$ foliating surfaces, along with the metric $\hat \gamma_{ij}$ induced on these level surfaces, the  primary metric $h_{ij}$ on $\Sigma$ can be decomposed as 
\begin{equation}\label{hij}
h_{ij}=\hat \gamma_{ij}+\hat  n_i \hat n_j\,,
\end{equation}
while the extrinsic curvature $\hat K_{ij}$ of $\mycal{S}_\rho$ can be given as
\begin{equation}\label{hatextcurv}
\hat K_{ij}= {{\hat \gamma}^{l}}{}_{i}\, D_l\,\hat n_j=\tfrac12\,\mycal{L}_{\hat n} {\hat \gamma}_{ij}\,.
\end{equation}

\medskip

Choose now a vector field $\rho^i$ satisfying the relation $\rho^i D_i \rho=1$ throughout $\Sigma$. The unit norm vector field $\hat n^i$ may also be decomposed as 
\begin{equation}\label{nhat}
\hat n^i={\hat{N}}^{-1}[\, \rho^i-{\hat N}{}^i\,]\,,
\end{equation}
where the `lapse' and `shift' of $\rho^i$ are determined via the relations $\hat n_i= \hat N (d\rho)_i$ and $\hat N^i=\hat \gamma{}^i{}_j\,\rho^j$, with $\hat\gamma{}^i{}_j=\delta{}^i{}_j-\hat n{}^i\hat n_j$. 

\medskip

In addition to the smooth function $\rho: \Sigma \rightarrow \mathbb{R}$ local coordinates  $(\rho,x^2,\dots,x^{n})$, adopted to the foliation $\mycal{S}_{\rho}$ and the vector field $\rho^i$, can also be defined\,\footnote{In these local coordinates the spatial indices of the pull backs of tensorial expressions to the $\mycal{S}_{\rho}$ level surfaces will be indicated by uppercase Latin indices and they always take the values $2,3,\dots,n$.} on subsets of $\Sigma$. This is done by choosing first (local) coordinates $x^A$ (with $A=2,3,\dots,n$) on patches of one of the level surfaces $\mycal{S}_{\rho_0}$ of the foliation of $\Sigma$ and then by Lie dragging them along the integral curves of $\rho^i$.

\medskip

The primary extrinsic curvature $K_{ij}$ can also be decomposed as 
\begin{equation}\label{decom_K}
K_{ij}= \boldsymbol\kappa \,\hat n_i \hat n_j  + \left[\hat n_i \,{\rm\bf k}{}_j  + \hat n_j\,{\rm\bf k}{}_i\right]  + {\rm\bf K}_{ij}\,,
\end{equation}
where 
\begin{equation}
\boldsymbol\kappa= \hat n^k\hat  n^l\,K_{kl},\ {\rm\bf k}{}_{i} = {\hat \gamma}^{k}{}_{i} \,\hat  n^l\, K_{kl}\ \ {\rm and} \ \ {\rm\bf K}_{ij} = {\hat \gamma}^{k}{}_{i} {\hat \gamma}^{l}{}_{j}\,K_{kl}\,.
\end{equation} 
Notice that all the boldfaced quantities may be considered as if they lived on the $\rho=const$ surfaces. 
We shall refer to the traces 
\begin{equation}
{\hat K}{}^l{}_{l}=\hat\gamma^{kl}\,{\hat K}_{kl}\ \ {\rm and} \ \ {\rm\bf K}^l{}_{l}=\hat\gamma^{kl}\,{\rm\bf K}_{kl}\,, 
\end{equation}
and also to the trace free part of ${\rm\bf K}_{ij}$ defined as 
\begin{equation}
\interior{\rm\bf K}_{ij}={\rm\bf K}_{ij}-\tfrac1{n-1}\,\hat \gamma_{ij}\,{\rm\bf K}^l{}_{l}\,. 
\end{equation}

Note finally that the $n\,(n+1)$ independent components of the original pair of variables $(h_{ij},K_{ij})$ may then be equivalently represented by the components of the variables in the septuple $(\hat N,\hat N^i,\hat \gamma_{ij}; \boldsymbol\kappa, {\rm\bf k}{}_{i}, {\rm\bf K}^l{}_{l},\interior{\rm\bf K}_{ij})$.  

\section{The constraints in new dress}\label{constraints} 
\setcounter{equation}{0}

By making use of the above decompositions, along with a sensible adaptation of (A.9)-(A.10) and (A.12)-(A.13) of \cite{racz_geom_det}, the momentum constraint can be seen to be equivalent to the system   
\begin{align}
\mycal{L}_{\hat n} {\rm\bf k}{}_{i} - \tfrac{n-2}{n-1}\,\hat D_i ({\rm\bf K}^l{}_{l}) - \hat D_i\boldsymbol\kappa + \hat D^l \interior{\rm\bf K}{}_{li} + ({\hat K^{l}}{}_{l})\,{\rm\bf k}{}_{i}  + \boldsymbol\kappa\,\dot{\hat n}{}_i - \dot{\hat n}{}^l\,{\rm\bf K}_{li} 
  -\epsilon\,\mathfrak{p}_l\,{\hat \gamma^{l}}{}_{i} = {}& 0 \label{par_const_n} \\
\mycal{L}_{\hat n}({\rm\bf K}^l{}_{l}) - \hat D^l {\rm\bf k}_{l} - \boldsymbol\kappa\,({\hat K^{l}}{}_{l})  + {\rm\bf K}{}_{kl}{\hat K}{}^{kl}  + 2\,\dot{\hat n}{}^l\, {\rm\bf k}_{l}  + \epsilon\,\mathfrak{p}_l\,{\hat n^{l}} = {}& 0\,, \label{ort_const_n}
\end{align}
where $\hat D_i$ denotes the covariant derivative operator associated with $\hat \gamma_{ij}$, and $\dot{\hat n}{}_k={\hat n}{}^lD_l{\hat n}{}_k=-{\hat D}_k(\ln{\hat N})$. 

\medskip

Notice that only the source terms---the last term on the l.h.s.~of (\ref{par_const_n}) and (\ref{ort_const_n}), respectively---pick up an $\epsilon$ factor which means that the principal parts of the above equations are not affected by the signature of the metric of the primary space. 

Remarkably, $\tfrac{n-1}{n-2}\,{\hat N}\,\hat \gamma{}^{ij} $ times of (\ref{par_const_n}) and ${\hat N}$ times of (\ref{ort_const_n}) when writing them out in (local) coordinates $(\rho,x^2,\dots,x^{n})$, adopted to the foliation $\mycal{S}_{\rho}$ and the vector field $\rho^i$, can be seen to take the form 
\begin{equation}\label{constr_hyp}
\left\{\hskip-.09cm \left(\hskip-.09cm
\begin{array}{cc}
 \hskip-.09cm \frac{n-1}{n-2}\,{\hat \gamma}{}^{AB} & \hskip-.19cm 0 \\ 
 \hskip-.19cm 0  &\hskip-.19cm 1
\end{array} 
\hskip-.09cm 
\right)
\hskip-.06cm
\,\partial_\rho +
\left(\hskip-.09cm
\begin{array}{cc}
 \hskip-.19cm -  \frac{n-1}{n-2}\,\hat N^{K}\,{\hat \gamma}{}^{AB} & \hskip-.19cm- {\hat N}\,{\hat \gamma}{}^{AK}  \\
 \hskip-.19cm - {\hat N}\,{\hat \gamma}{}^{BK} &  \hskip-.19cm -  \hat N^{K}
\end{array} \hskip-.2cm
\right)\hskip-.06cm
\,\partial_K\hskip-.06cm
\right\}
\left(\hskip-.19cm
\begin{array}{c}
{\rm\bf k}_{B} \\
\\
{\rm\bf K}^E{\hskip-.5mm}_{E}
\end{array} \hskip-.19cm
\right) +
\left(\hskip-.19cm
\begin{array}{c}
\mycal{B}^{\,A}_{({\rm\bf k})} \\
\mycal{B}_{({\rm\bf K})}
\end{array} \hskip-.19cm
\right)=0 \,.
\end{equation}
Notice that the coefficients of $\partial_\rho$ and $\partial_K$ are symmetric, and that the coefficient of $\partial_\rho$ is also positive definite. Thereby (\ref{constr_hyp}) possesses the form of a first order symmetric hyperbolic system 
\begin{equation}\label{constr_hyp2}
\mathcal A^{(\rho)} \,\partial_\rho {\bf u} + \mathcal A^{(K)} \,\partial_K {\bf u}+ \mathcal B = 0 
\end{equation}
(with $K=2,\dots,n$) with the vector valued variable 
\begin{equation}\label{vector-valued}
{\bf u}=({\rm\bf k}_{B}, {\rm\bf K}^E{\hskip-.5mm}_{E})^T\,,
\end{equation}
and such that the `radial coordinate' $\rho$ plays the role of `time'.

\medskip

Notice also that the characteristic cone of (\ref{constr_hyp})---apart from the $\mycal{S}_{\rho}$ codimension-one surfaces ($\hat n^i \xi_i = 0$ on them) foliating $\Sigma$---reads as 
\begin{equation}\label{char_cone}
[{\hat \gamma}^{ij} - (n-1)\, \hat n^i \hat n^j]\, \xi_i \xi_j =  0\,.
\end{equation}
It is also remarkable that all the coefficients and the source terms in (\ref{constr_hyp}) may be considered as if they were defined on the level surfaces $\mycal{S}_{\rho}$ exclusively. 


\bigskip

In turning to the Hamiltonian constraint recall first that in virtue of (A.1) of \cite{racz_geom_det} for the scalar curvature\,\footnote{Here, as the metric $h_{ij}$ on $\Sigma$ was assumed to be Riemannian, regardless whether the metric of the primary space was of Euclidean or Lorentzian signature, the value of $\epsilon$---that applies in a sensible adaptation of (A.1) of \cite{racz_geom_det}---is $+1$.} of $h_{ij}$ 
\begin{equation}
{}^{{}^{(n)}}\hskip-1mm R= \hat R - \left\{2\,\mycal{L}_{\hat n} ({\hat K^l}{}_{l}) + ({\hat K^{l}}{}_{l})^2 + \hat K_{kl} \hat K^{kl} + 2\,{\hat N}^{-1}\,\hat D^l \hat D_l \hat N \right\}
\end{equation}
holds, where $\hat R$ denotes the scalar curvature associated with $\hat \gamma_{ij}$. Combining this with the decomposition (\ref{decom_K}) of the primary extrinsic curvature $K_{ij}$ the Hamiltonian constraint (\ref{expl_eh}) can be put into the form 
\begin{align}
-\epsilon\,\hat R + \epsilon\left\{2\,\mycal{L}_{\hat n} ({\hat K^l}{}_{l})\right.\hskip-1mm {}& {} \left.  + \hskip1mm ({\hat K^{l}}{}_{l})^2 + {\hat K}{}_{kl}\,{\hat K}{}^{kl} +  2\,{\hat N}^{-1}\,\hat D^l \hat D_l \hat N \right\} \nonumber \\ {}&  + 2\,\boldsymbol\kappa\,({\rm\bf K}^l{}_{l})+\tfrac{n-2}{n-1}\,({\rm\bf K}^l{}_{l})^2 
-2\,{\rm\bf k}{}^{l}{\rm\bf k}{}_{l}  - \interior{\rm\bf K}{}_{kl}\,\interior{\rm\bf K}{}^{kl} -2\,\mathfrak{e}=0\,. \label{scal_constr_n}
\end{align}

\section{The evolutionary systems}\label{constraints}
\setcounter{equation}{0}

The system comprised by (\ref{par_const_n})-(\ref{ort_const_n}) and (\ref{scal_constr_n}) has to be solved simultaneously. This section is to discuss two cases---each corresponding to a specific choice for the controlled variables in (\ref{scal_constr_n})---such that both of these lead to an evolutionary setup. 
In particular, if (\ref{scal_constr_n}) is solved for the lapse $\hat N$ the corresponding equation becomes parabolic, and the coupled equations (\ref{par_const_n})-(\ref{ort_const_n}) and (\ref{scal_constr_n}) determine then a hyperbolic-parabolic system to which an initial value problem can be setup. Alternatively, (\ref{scal_constr_n}) can also be solved algebraically for $\boldsymbol\kappa$, and the corresponding set of equations (\ref{par_const_n})-(\ref{ort_const_n}) and (\ref{scal_constr_n}) turn to be a hyperbolic-algebraic system. Notably, by eliminating  $\boldsymbol\kappa$ and its derivatives from (\ref{par_const_n})-(\ref{ort_const_n}), under suitable conditions, a  Friedrichs symmetrizable hyperbolic system can be deduced from them to which a well-posed initial value problem known to apply. 

\medskip

The following subsections are to deduce these hyperbolic-parabolic and symmetrizable hyperbolic systems and to identify those conditions which guarantee the existence and uniqueness of local solutions to the pertinent equations.  

\subsection{The hyperbolic-parabolic system}

This subsection is to show that if (\ref{scal_constr_n}) is solved for the lapse $\hat N$, and if the $\mycal{S}_{\rho}$ level surfaces of the foliation of $\Sigma$ are of positive mean curvature, a parabolic problem manifest itself, and that the coupled equations (\ref{par_const_n})-(\ref{ort_const_n}) and (\ref{scal_constr_n}) give rise to a locally well-posed hyperbolic-parabolic system. 

\medskip

To see that this is indeed the case recall first that 
\begin{equation}\label{hatext}
\hat K_{ij} = \tfrac12\,\mycal{L}_{\hat n} {\hat \gamma}_{ij} = \hat N{}^{-1}[\,\tfrac12\mycal{L}_{\rho} {\hat \gamma}_{ij} -\hat  D_{(i}\hat N_{j)}\,] = \hat N{}^{-1}\instar{K}_{ij}\,,
\end{equation}
and that 
\begin{equation}\label{trhatext}
\hat K{}^l{}_{l} = {\hat \gamma}^{ij}\,\hat K_{ij} = \hat N{}^{-1}[\,\tfrac12\,{\hat \gamma}^{ij}\mycal{L}_{\rho} {\hat \gamma}_{ij} -  \hat D_j\hat N^j\,]= \hat N{}^{-1}\instar{K} \,,
\end{equation}
where $\instar{K}_{ij}$ and $\instar{K}={\hat \gamma}^{ij}\instar{K}_{ij}$ were introduced only for notational conveniences. 
It is also straightforward to see that 
\begin{equation}\label{lietrhatext}
\mycal{L}_{\hat n} (\hat K{}^l{}_{l}) = - \hat N{}^{-3}\instar{K} \,[\,(\partial_{\rho} \hat N) - (\hat N{}^l\hat D_l\hat N)\,] + \hat N{}^{-2} [\,(\partial_{\rho} \instar{K}) - (\hat N{}^l \hat D_l \instar{K})\,] \,.
\end{equation}
By substituting the above relations into (\ref{scal_constr_n}) we get 
\begin{align}
-\epsilon\,\hat R + \epsilon\left\{ - 2\,\hat N{}^{-3}\instar{K} [\,(\partial_{\rho} \hat N) - (\hat N{}^l\hat D_l\hat N)\,] \right.\hskip-1mm {}& {} \left. + \,2\,\hat N{}^{-2} \,[\,(\partial_{\rho} \instar{K}) - (\hat N{}^l \hat D_l \instar{K})\,] \right. \nonumber \\ {}& \left. +\, \hat N{}^{-2} \,\instar{K}^2 + \hat N{}^{-2} \,\instar{K}{}_{kl} \instar{K}{}^{kl} +  2\,{\hat N}^{-1}\hat D^l \hat D_l \hat N \right\} \nonumber \\ {}& \hskip-3cm + 2\,\boldsymbol\kappa\,({\rm\bf K}^l{}_{l})+\tfrac{n-2}{n-1}\,({\rm\bf K}^l{}_{l})^2 
-2\,{\rm\bf k}{}^{l}{\rm\bf k}{}_{l}  - \interior{\rm\bf K}{}_{kl}\,\interior{\rm\bf K}{}^{kl} -2\,\mathfrak{e}=0\,, \label{par_scal_constr_n}
\end{align}
which by applying the notation 
\begin{align}
A = {}& 2\,[\,(\partial_{\rho} \instar{K}) - \hat N{}^l (\hat D_l \instar{K})\,] + \instar{K}^2 + \instar{K}{}_{kl} \instar{K}{}^{kl} \label{A}  \\
B = {}& - \hat R + \epsilon\,\bigl[\, 2\,\boldsymbol\kappa\,({\rm\bf K}^l{}_{l})+\tfrac{n-2}{n-1}\,({\rm\bf K}^l{}_{l})^2 
-2\,{\rm\bf k}{}^{l}{\rm\bf k}{}_{l}  - \interior{\rm\bf K}{}_{kl}\,\interior{\rm\bf K}{}^{kl} -2\,\mathfrak{e}\,\bigr] \label{B}
\end{align}
can be put into the form 
\begin{equation}\label{bern_pde}
2\,\instar{K}\,[\,(\partial_{\rho} \hat N) - \hat N{}^l(\hat D_l\hat N) \,] = 2\,\hat N^{2} (\hat D^l \hat D_l \hat N) + A\,\hat N + B\,\hat N{}^{3}\,,
\end{equation}
which is a Bernoulli-type parabolic partial differential equation provided that $\instar{K}$, or equivalently---as, it will be shown below, $\hat N$ can be guaranteed to be positive---$\hat K{}^l{}_{l}=\hat N{}^{-1}\instar{K}$ is positive throughout $\Sigma$, i.e.~$\mycal{S}_\rho$ is a positive mean curvature foliation of $\Sigma$. Accordingly, if this happens equation (\ref{bern_pde}) is uniformly parabolic such that $\rho$ plays the role of ``time''. 

\medskip

Note that, in virtue of (\ref{trhatext}), 
\begin{equation}\label{Kstar}
\instar{K}=\tfrac12\,{\hat \gamma}^{ij}\mycal{L}_{\rho} {\hat \gamma}_{ij} -  \hat D_j\hat N^j\,,
\end{equation}
and that both terms on the right hand side can be evaluated once $\hat N^i$ and $\hat \gamma_{ij}$ are known on $\Sigma$. Since these variables will be seen to belong to the freely specifiable part of the data, as listed in (\ref{list-free_a}), they can be chosen so that $\instar{K}$ is guaranteed to be positive throughout $\Sigma$. This verifies then that equation (\ref{bern_pde}) is indeed uniformly parabolic. If, in addition, smooth positive initial data ${}_0\hat N$ is given on a level surface $\rho=\rho_0$ of the foliation of $\Sigma$, then standard results guarantee the existence of a unique positive solution for $\hat N$, with $\hat N\vert_{\mycal{S}_{\rho_0}}={}_0\hat N$, on a short interval $[\rho_0,\rho_0+\varepsilon)$ for some $\varepsilon>0$, i.e.~in a one-sided neighborhood $\mycal{S}_{[\rho_0,\rho_0+\varepsilon)}$ of $\mycal{S}_{\rho_0}$ in $\Sigma$. 

\medskip

Note that as the fundamental characters of the principal parts of (\ref{par_const_n})-(\ref{ort_const_n}) and (\ref{scal_constr_n}), respectively, is not affected by the coupling of these first order symmetric hyperbolic and parabolic systems the following short time existence and uniqueness results apply to this coupled system:

\begin{theorem}
Suppose that a smooth choice had been made for the freely specifiable variables
\begin{equation}\label{list-free_a}
\hat N^i,\hat \gamma_{ij}\,; \boldsymbol\kappa, \interior{\rm\bf K}_{ij}\,; \mathfrak{e}, \mathfrak{p}_i
\end{equation}
so that $\instar{K}>0$ throughout $\Sigma$. Assume that smooth initial data $({}_0\hat N, {}_0{\rm\bf k}_{i}, {}_0{\rm\bf K}^l{ }_{l})$, with ${}_0\hat N>0$, had also been chosen to the hyperbolic-parabolic system on one of the level surfaces $\mycal{S}_{\rho_0}$ in $\Sigma$. Then, on a short interval $[\rho_0,\rho_0+\varepsilon)$, for some $\varepsilon>0$, there exists a unique smooth solution 
$(\hat N, {\rm\bf k}_{i}, {\rm\bf K}^l{ }_{l})$, with $\hat N>0$,
to the hyperbolic-parabolic system comprised by (\ref{par_const_n})-(\ref{ort_const_n}) and (\ref{scal_constr_n}) such that $\hat N\vert_{\mycal{S}_{\rho_0}}={}_0\hat N, {\rm\bf k}_{i}\vert_{\mycal{S}_{\rho_0}}={}_0{\rm\bf k}_{i}, {\rm\bf K}^l{ }_{l}\vert_{\mycal{S}_{\rho_0}}= {}_0{\rm\bf K}^l{ }_{l}$. The fields $h_{ij}$ and $K_{ij}$ which can be built up from this solution and the freely specifiable part of the data satisfy the full constraint system (\ref{expl_eh})-(\ref{expl_em}) in a one-sided neighborhood $\mycal{S}_{[\rho_0,\rho_0+\varepsilon)}$ of $\mycal{S}_{\rho_0}$ in $\Sigma$.
\end{theorem} 

\medskip

Clearly, it would be of obvious interest to know whether by imposing certain conditions on the coefficients and on the initial data the existence of unique solutions to this coupled hyperbolic-parabolic  on the global time interval $[\rho_0,\infty)$ could also be shown. It would be even more important to identify those conditions which could guarantee that the initial data $(h_{ij},K_{ij})$---that could be recovered from these solutions and from the freely specifiable part of the data---was of asymptotically Euclidean. The study of all these issues is out of the scope of the present paper and they are left for future investigations. However, as for certain choice of coefficients in (\ref{bern_pde}) and the initial data ${}_0\hat N$ on ${\mycal{S}_{\rho_0}}$ classical solutions are known to blow up in finite ``time'' it is of obvious interest to know those conditions which guarantee the existence of global in the time, $[\rho_0,\infty)$, solutions to (\ref{bern_pde}).  

\medskip

In identifying these conditions we shall apply an argument adapted from Bartnik \cite{bartnik} (see also \cite{smithwein1,smithwein2}). Notably many details of Bartnik's  proofs developed to study the existence of quasi-spherical foliations in asymptotically flat spacetimes do also apply in the generic setup used in this paper. For instance, consider a function $\tf$ on $\Sigma$ and, following Bartnik's notation in \cite{bartnik}, define $\tf^*$ and $\tf_*$ as
\begin{equation}
\tf^*(\rho)=\sup_{\mycal{S}_{\rho}} \,\{\,\tf(\rho,x^2,\dots,x^n) \,\} \,,\quad \tf_*(\rho)=\inf_{\mycal{S}_{\rho}} \,\{\, \tf(\rho,x^2,\dots,x^n) \,\}\,.
\end{equation}
Substituting then $\hat N$ in (\ref{bern_pde}) by applying the relation $w=\hat N{}^{-2}$ we get  
\begin{equation}\label{bern_pde_2}
\partial_{\rho} w + \hat N{}^{-3}\hat N{}^l(\hat D_l \hat N) = -2\,\instar{K}^{-1}\hat N{}^{-1} (\hat D^l \hat D_l \hat N) - a\,w - b\,,
\end{equation}
where $a=A/\instar{K}$ and $b=B/\instar{K}$. Notice that in the ``gradient'' and ``Laplacian'' terms the substitution was not performed as  the parabolic maximum principle, along with the fact that $\hat N$ has a {\it maximum} where $w$ has a {\it minimum} and, similarly, $\hat N$ has a {\it minimum} where $w$ has a {\it maximum}, implies that at the maximum of $\hat N$ the relation
\begin{equation}
\partial_{\rho} w_*\geq - a^*\,w_* - b^*
\end{equation}
holds. This immediately gives us a lower bound
\begin{equation}\label{w^*lower}
 w_* \geq W_{*} = 
\exp\left[-\int_{\rho_0}^\rho a^* \d\rho' \right] \left\{ w_*\vert_{\mycal{S}_{\rho_0}}  -  \int_{\rho_0}^\rho b^* \exp\left[\int_{\rho_0}^{\rho'} a^* \d\rho'' \right]\d\rho'\right\} \,.
\end{equation} 
By a completely analogous argument we also get an upper bound 
\begin{equation}\label{w^*upper}
 w^* \leq W^* = 
\exp\left[-\int_{\rho_0}^\rho a_* \d\rho' \right] \left\{ w^*\vert_{\mycal{S}_{\rho_0}}  -  \int_{\rho_0}^\rho b_* \exp\left[\int_{\rho_0}^{\rho'} a_* \d\rho'' \right]\d\rho'\right\} \,.
\end{equation}
Based on (\ref{w^*upper}), and following the argument in \cite{smithwein1}, we may conclude, on the one hand side, that $w$ is bounded from above, $w\leq W^*$, which implies that $\hat N$ is positive as then $0 < (W^*)^{-\tfrac12} \leq w^{-\tfrac12}=\hat N$. If, in addition, 
\begin{equation}
\mathcal{K} = 
\sup_{\rho\in[\rho_0,\infty)} \left\{\int_{\rho_0}^\rho b^* \exp\left[\int_{\rho_0}^{\rho'} a^* \d\rho'' \right]\d\rho' \right\}  
\end{equation}
is positive and finite, i.e.~$0<\mathcal{K}<\infty$ and if for the initial data $w_*\vert_{\mycal{S}_{\rho_0}}$ the inequality $w_*\vert_{\mycal{S}_{\rho_0}}>\mathcal{K}$ holds then it follows from (\ref{w^*lower}) that $w$ is also bounded from below by $W_{*}>0$. This implies, in turn, that $\hat N$ has to be bounded, i.e.~$\hat N = w^{-\tfrac12}\leq W_{*}{}^{-\tfrac12} <  \infty$ provided that for the initial data ${}_0\hat N$ the inequality ${}_0\hat N^{-2}=w\vert_{\mycal{S}_{\rho_0}}\geq w_*\vert_{\mycal{S}_{\rho_0}}>\mathcal{K}$, or equivalently ${}_0\hat N<1/\sqrt{\mathcal{K}}$, holds. Remarkably, if $B$, given by (\ref{B}), is negative then $\mathcal{K}<0$ from which it follows that for any positive choice of the initial data ${}_0\hat N$ to (\ref{bern_pde}) $W_{*}$ is positive and, in turn, the relation $\hat N<\infty$ follows. 

\medskip

By combining all the above observations we get

\begin{theorem}\label{bern_pde_theor}
Suppose that all the coefficients in (\ref{bern_pde}) are smooth and that the freely specifiable part of the data was chosen such that $\instar{K}$ is positive throughout $\Sigma$. Choose ${}_0\hat N$ to be a smooth positive function on the level surface $\rho=\rho_0$ such that ${}_0\hat N<1/\sqrt{\mathcal{K}}$ if $0<\mathcal{K}<\infty$, or to be arbitrary if $\mathcal{K}\leq 0$. Then  (\ref{bern_pde}) has a unique smooth global in the time, $[\rho_0,\infty)$, classical solution such that $0<\hat N<\infty$, and that $\hat N\vert_{\mycal{S}_{\rho_0}} = {}_0\hat N$. 
\end{theorem}

\medskip

We close this subsection by comparing the assumptions used in our setup and those applied by Bartnik's in \cite{bartnik} and by Smith and Weinstein in \cite{smithwein1,smithwein2}. To do so recall first that in all of these works, as opposed to the setup proposed in this paper, attention was restricted exclusively to the time symmetric case with vanishing $K_{ij}$. 
Furthermore, in \cite{bartnik,smithwein1,smithwein2} the foliating $\mycal{S}_{\rho}$ surfaces were assumed to be topological two-spheres whereas in our setup they were allowed to be generic as far as their mean curvature is positive. Moreover, the metric $\hat \gamma_{ij}$ in Bartnik's quasi-spherical foliations was supposed to possess the form $\hat \gamma_{ij}=r^2\,\interior{\gamma}_{ij}$, where  $\interior{\gamma}_{ij}$ stands for the unit two sphere metric and the function $r:\Sigma \rightarrow \mathbb{R}$ determines the level surfaces of the foliation of $\Sigma$. This setup was slightly generalized by Smith and Weinstein in \cite{smithwein1,smithwein2} by assuming that $\hat \gamma_{ij}$ to possess the form $\hat \gamma_{ij}=e^{2v}r^2\,\interior{\gamma}_{ij}$ with some smooth function $v:\Sigma \rightarrow \mathbb{R}$ though they also required that the $r=conts$ level surfaces have area $4\pi r^2$. Both of these foliations are quasi-convex in the sense that the Gauss and mean curvatures of the level 
surfaces of the foliation $\mycal{S}_{\rho}$ are positive, and the quasi-spherical foliation of Bartnik can be recovered by setting $v=0$ in \cite{smithwein1,smithwein2}.

\subsection{The symmetrizable hyperbolic system}

As mentioned above (\ref{scal_constr_n}) may also be solved for $\boldsymbol\kappa$.  Since (\ref{par_const_n}) involves the ``angular'' derivatives of $\boldsymbol\kappa$ as a first step in demonstrating the viability of the proposed method is to eliminate these derivatives and then to show the solvability of the yielded system. 

\medskip

Once the derivatives $\hat D_i\boldsymbol\kappa$ are eliminated from  (\ref{par_const_n}) the reduced system can be seen to be comprised by  (\ref{ort_const_n}) and
\begin{align}\label{constr_mom1_n} 
\mycal{L}_{\hat n} {\rm\bf k}{}_{i}  + ({\rm\bf K}^l{}_{l})^{-1}[\,\boldsymbol\kappa\,\hat D_i ({\rm\bf K}^l{}_{l}) -2\, {\rm\bf k}{}^{l}\hat D_i{\rm\bf k}{}_{l}\,] + (2\,{\rm\bf K}^l{}_{l})^{-1}\hat D_i\boldsymbol\kappa_0 {}& \\ 
+ ({\hat K^{l}}{}_{l})\,{\rm\bf k}{}_{i} + [\,\boldsymbol\kappa-\tfrac1{n-1}\, ({\rm\bf K}^l{}_{l})\,]\,\dot{\hat n}{}_i  - \dot{\hat n}{}^l\,\interior{\rm\bf K}_{li}  +{}&  \hat D^l \interior{\rm\bf K}{}_{li}  -\epsilon\,\mathfrak{p}_l\,{\hat \gamma^{l}}{}_{i} = 0\,, \nonumber  
\end{align}
where $\boldsymbol\kappa$ and $\boldsymbol\kappa_0$ are supposed to be replaced by the algebraic expressions 
\begin{equation} \label{constr_ham_n} 
\boldsymbol\kappa= (2\,{\rm\bf K}^l{}_{l})^{-1}[\,2\,{\rm\bf k}{}^{l}{\rm\bf k}{}_{l} - \tfrac{n-2}{n-1}\,({\rm\bf K}^l{}_{l})^2 - \boldsymbol\kappa_0\,]
\end{equation}
\begin{equation} \label{kappa0} 
\boldsymbol\kappa_0= -\epsilon\,\hskip-1mm {}^{{}^{(n)}}\hskip-1mm R - \interior{\rm\bf K}{}_{kl}\,\interior{\rm\bf K}{}^{kl} -2\,\mathfrak{e}\,.
\end{equation}
Note that (\ref{constr_ham_n}), with the substitution of (\ref{kappa0}), represents the algebraic solution of (\ref{scal_constr_n}) for $\boldsymbol\kappa$, and that $\boldsymbol\kappa_0$ depends only on the quantities $\hat N,\hat N^i,\hat \gamma_{ij}$ and $\interior{\rm\bf K}_{ij}$ which, in the present case, will be seen to comprise the freely specifiable part of the initial on $\Sigma$. 

\medskip

Note that the above outlined process guarantee that solutions $\boldsymbol\kappa, {\rm\bf k}{}_{i}, {\rm\bf K}^l{}_{l}$ to the system comprised by (\ref{constr_mom1_n}) and (\ref{ort_const_n}), along with the freely specifiable part of the data  $\hat N,\hat N^i,\hat \gamma_{ij}, \interior{\rm\bf K}_{ij}$, in virtue of the relations (\ref{hij}) and (\ref{decom_K}) determine solutions to the full constraint system (\ref{expl_eh})-(\ref{expl_em}). 

\medskip

Notably, when writing out (\ref{constr_mom1_n}) and (\ref{ort_const_n}) in local coordinates $(\rho,x^2,\dots,x^{n})$, adopted to the foliation $\mycal{S}_{\rho}$ and the vector field $\rho^i$, they take the form 
\begin{equation}\label{constr_hyp2}
\partial_\rho {\bf u} + \mathcal B^{(K)} \,\partial_K {\bf u}+ \mathcal C = 0 
\end{equation}
for the vector valued variable 
${\bf u}=({\rm\bf k}_{B}, {\rm\bf K}^E{\hskip-.5mm}_{E})^T$,
where the coefficients $\mathcal B^{(K)}$ of $\partial_K$, with $K=2,3,\dots,n$, can be given as 
\begin{equation}\label{constr_hyp2_coeffs}
\mathcal B^{(K)}_{\,ij}=- \hat N{}^K\delta_{ij} - \hat N{}\,\hat\gamma^{KL}\,\delta_{in}\,\delta_{j(L-1)} + \frac{\boldsymbol\kappa\,\hat N{}}{{\rm\bf K}^E{\hskip-.5mm}_{E}} \,\delta_{i(K-1)}\,\delta_{jn} - 2\,\frac{\hat N{}\,{\rm\bf k}_{L}}{{\rm\bf K}^E{\hskip-.5mm}_{E}}\,\hat\gamma^{LM}\,\delta_{i(K-1)}\,\delta_{(j+1)M} \,,
\end{equation}
and where the lowercase and uppercase Latin indices take the value $1,2,\dots,n$ and $2,3,\dots,n$, respectively.\footnote{As the lowercase and uppercase Latin indices take the value $1,2,\dots,n$ and $2,3,\dots,n$, respectively, in evaluating a term of the form $\delta_{(j+1)M}$, here and also in (\ref{constr_hyp2_coeffs_symmetric}), one needs to refer to the $(n+1)$-dimensional Kronecker delta.} 

\medskip

A system of the form of (\ref{constr_hyp2}) is known to be Friedrichs symmetrizable (see, e.g.~\cite{benzoni,kreissl,reula}) if a common symmetrizer of the coefficients $\mathcal B^{(K)}$ exists. A bilinear form, $\mathcal H_{ij}$, is called to be a common symmetrizer to (\ref{constr_hyp2}) if it is positive definite and is such that for any value of $K=2,\dots,n$ the product ${\mathcal H}_{ij}\,\mathcal B^{(K)}_{\,jk}$ is symmetric. It may be verified by a direct calculation that the matrix of the form
\begin{align}\label{constr_hyp2_coeffs_symmetric}
\mathcal H_{\,ij}=\delta_{in}\,\delta_{jn} {} & -\frac{2}{\boldsymbol\kappa} \,{\rm\bf k}_{P}\,\hat\gamma^{PQ}\left[\,\delta_{in}\,\delta_{(j+1)Q}+\delta_{(i+1)Q}\,\delta_{jn} \,\right] \\  {} & + \frac{1}{{\boldsymbol\kappa}^2}\left[\,4\,({\rm\bf k}_{P}\,\hat\gamma^{PQ}\,\delta_{(i+1)Q})\,({\rm\bf k}_{S}\,\hat\gamma^{ST}\,\delta_{(j+1)T}) - \boldsymbol\kappa\cdot{{\rm\bf K}^E{\hskip-.5mm}_{E}}\,\hat\gamma^{PQ}\,\delta_{(i+1)P}\,\delta_{(j+1)Q} \,\right]  \nonumber
\end{align}
is indeed a common symmetrizer to (\ref{constr_hyp2}) in a subset of $\Sigma$ where neither $\boldsymbol\kappa$ nor ${\rm\bf K}^l{}_{l}$ vanishes and in which they are of opposite sign. This observation is summarized in the following lemma.

\begin{lemma}
Assume that $\boldsymbol\kappa\cdot{\rm\bf K}^l{}_{l}<0$ holds in a subset of $\Sigma$. Then equations (\ref{constr_mom1_n}) and (\ref{ort_const_n}) comprise a first order symmetrizable hyperbolic system there.
\end{lemma}

\medskip

It is also informative to inspect the characteristic directions $\xi_i=(\xi_\rho, \xi_2,\dots,\xi_n)$ which are the roots of the essential part of the characteristic polynomial
\begin{equation}\label{char_cone_2}
{\rm\bf K}^l{}_{l}\,(\hat n^i\xi_i)^2 - 2\,(\hat n^i\xi_i)\,({\hat \gamma}^{KL}{\rm\bf k}{}_K \xi_L) +\boldsymbol\kappa\,({\hat \gamma}^{KL}\xi_K \xi_L)   =  0
\end{equation}
of (\ref{constr_hyp2}). In virtue of (\ref{nhat}) and (\ref{char_cone_2}), for the $\rho$-component of a characteristic direction the relation 
\begin{align}\label{xirho}
\xi_\rho=  \hat N{}^K\xi_K + \frac{\hat N{}}{{\rm\bf K}^E{\hskip-.5mm}_{E}}\left[({\hat \gamma}^{KL}{\rm\bf k}{}_K \xi_L) \pm \sqrt{({\hat \gamma}^{KL}{\rm\bf k}{}_K \xi_L)^2-\boldsymbol\kappa\cdot{{\rm\bf K}^E{\hskip-.5mm}_{E}}\,({\hat \gamma}^{KL}\xi_K \xi_L)} \,\right]
\end{align}
holds. Notice that in (\ref{xirho}) the inequality $\boldsymbol\kappa\cdot{\rm\bf K}^l{}_{l}<0$ comes into play again.

\medskip

The above discussion indicates that it is of crucial importance to know if this sign condition may be guaranteed to hold. 

\medskip

To get some hints about the viability of the proposed setup note first that, in virtue of (\ref{hij}) and (\ref{decom_K}), the relation 
\begin{equation}\label{decom_K_2}
K^e{}_{e}= \boldsymbol\kappa + {\rm\bf K}^l{}_{l}
\end{equation}
holds. It follows then that for any maximal ($K^e{}_{e}=0$) but non-time symmetric ($K{}_{ef}\not \equiv 0$ ) initial data surface the signs of $\boldsymbol\kappa$ and ${\rm\bf K}^l{}_{l}$ have to be opposite except where they both vanish simultaneously.  

\medskip

As it would be preferable to involve more generic initial data surfaces than the maximal ones it is important that by choosing the freely specifiable part of data on $\Sigma$, along with the initial data ${}_0{\rm\bf k}{}_{i}$ and ${}_0{\rm\bf K}^l{}_{l}$ for the system (\ref{constr_mom1_n}) and (\ref{ort_const_n}), properly the inequality $\boldsymbol\kappa\cdot{\rm\bf K}^l{}_{l}<0$ 
can always be guaranteed to hold locally. To see this note first that the initial data ${}_0{\rm\bf k}{}_{i}$ and ${}_0{\rm\bf K}^l{}_{l}$ for (\ref{constr_mom1_n}) and (\ref{ort_const_n}) is freely specifiable at one of the level surfaces of the foliation denoted by $\mycal{S}_{\rho_0}$. Using this freedom choose ${}_0{\rm\bf K}^l{}_{l}$ to be strictly non-zero there. This, implies then that for any choice of a continuous field ${\rm\bf K}^l{}_{l}$ on $\Sigma$ such that ${\rm\bf K}^l{}_{l}\vert_{\mycal{S}_{\rho_0}}={}_0{\rm\bf K}^l{}_{l}$ the field ${\rm\bf K}^l{}_{l}$ does not vanish in a sufficiently small neighborhood of 
$\mycal{S}_{\rho_0}$ 
either. Similarly, by choosing the initial data for ${}_0{\rm\bf k}{}_{i}$, along with the freely specifiable data on $\Sigma$, properly the term $\boldsymbol\kappa_0-\,2\,{\rm\bf k}{}^{l}{\rm\bf k}{}_{l}$ can also be guaranteed to be strictly positive on $\mycal{S}_{\rho_0}$, and, in turn, in a sufficiently small neighborhood of $\mycal{S}_{\rho_0}$ in $\Sigma$. 

\medskip

Viewing then (\ref{scal_constr_n}) as quadratic polynomial 
\begin{equation}\label{quadr_pol}
({\rm\bf K}^l{}_{l})^2+ \tfrac{2\,(n-1)}{n-2}\,\boldsymbol\kappa\, ({\rm\bf K}^l{}_{l}) + [\,\boldsymbol{\kappa}_\circ-\,2\,{\rm\bf k}{}^{l}{\rm\bf k}{}_{l} \,] =0 
\end{equation}
in ${\rm\bf K}^l{}_{l}$, and applying Vi${\grave{e}}$te's formula relevant for the product of the roots $({\rm\bf K}^l{}_{l})_1$ and $({\rm\bf K}^l{}_{l})_2$ we get 
\begin{equation} 
({\rm\bf K}^l{}_{l})_1({\rm\bf K}^l{}_{l})_2= \boldsymbol{\kappa}_0-\,2\,{\rm\bf k}{}^{l}{\rm\bf k}{}_{l}>0\,.
\end{equation}
This verifies then that the two roots are real and they are both either negative or positive in a neighborhood of $\mycal{S}_{\rho_0}$.  
In virtue of the other Vi${\grave{e}}$te's formula, pertinent for the sum of the roots, we have that
\begin{equation} 
({\rm\bf K}^l{}_{l})_1+({\rm\bf K}^l{}_{l})_2=-\tfrac{2\,(n-1)}{n-2}\,\boldsymbol{\kappa}\,,
\end{equation}
which implies then that $\boldsymbol\kappa$ and $({\rm\bf K}^l{}_{l})$ are of opposite sign at least in a sufficiently small neighborhood of $\mycal{S}_{\rho_0}$ in $\Sigma$. 

\medskip

As for symmetrizable hyperbolic systems the initial value problem is known to be well-posed in the smooth setting, in virtue of the above observations, we immediately have the following:  

\begin{theorem}
Assume that the freely specifiable part of data
\begin{equation}\label{list-free_b}
\hat N, \hat N^i,\hat \gamma_{ij}\,; \interior{\rm\bf K}_{ij}\,; \mathfrak{e}, \mathfrak{p}_i
\end{equation}
on $\Sigma$, and the initial data $({}_0{\rm\bf k}_{i}, {}_0{\rm\bf K}^l{ }_{l})$ for the system  (\ref{constr_mom1_n}) and (\ref{ort_const_n}) on a $\rho=\rho_0$ level surface are chosen such that they are all smooth on their respective domains, and that the system comprised by (\ref{constr_mom1_n}) and (\ref{ort_const_n}) can be guaranteed to be symmetrizable hyperbolic in a sufficiently small neighborhood of $\mycal{S}_{\rho_0}$ in $\Sigma$, as indicated above.  Then, there exists a unique smooth (local) solution 
$({\rm\bf k}_{i},{\rm\bf K}^l{ }_{l})$
to (\ref{constr_mom1_n}) and (\ref{ort_const_n}) such that ${\rm\bf k}_{i}\vert_{\mycal{S}_{\rho_0}}={}_0{\rm\bf k}_{i}, {\rm\bf K}^l{ }_{l}\vert_{\mycal{S}_{\rho_0}}= {}_0{\rm\bf K}^l{ }_{l}$. This solution, along with the freely specifiable data (\ref{list-free_b}), uniquely  determines $\boldsymbol{\kappa}$ via (\ref{constr_ham_n}). The fields $h_{ij}$ and $K_{ij}$ which can be built up from the solution to (\ref{constr_mom1_n}), (\ref{ort_const_n}) and  (\ref{constr_ham_n}) and from the freely specifiable part of the data do satisfy the full constraint system (\ref{expl_eh})-(\ref{expl_em}) in the domain of dependence of $\mycal{S}_{\rho_0}$ in $\Sigma$.
\end{theorem}

\medskip

To give an explicit example that have immediate physical relevance and where $\boldsymbol\kappa$ and ${\rm\bf K}^l{}_{l}$ satisfy globally the inequality $\boldsymbol\kappa\cdot{\rm\bf K}^l{}_{l}<0$ on some initial data surface restrict our considerations to the space of near Schwarzschild configurations. Recall that the Schwarzschild metric in the Kerr-Schild form on a Minkowski background $(\mathbb{R}^4, \eta_{ab})$ read as
\begin{equation} 
g_{ab}=\eta_{ab}+2 H \ell_a \ell_b\,,
\end{equation}
where $H=M/r$, which (apart from the singularity at the center, $r=0$) is smooth and $\ell_a$ is a null vector field with respect to both $g_{ab}$ and  $\eta_{ab}$. Choose now a $t=const$ Kerr-Schild time slice as our initial data surface $\Sigma$ in a generic Kerr-Schild setup, and assumed that the freely specifiable part of the data approximates that of the Schwarzschild solution and $\Sigma$ is foliated by ``$r=const$'' level surfaces. In particular, if in addition the relation $\frac{{\rm\bf k}{}_A}{\boldsymbol\kappa} \approx 0$ is satisfied on $\Sigma$, then, as it was shown in \cite{racz_winicour}, for these ``near Schwarzschild configurations'' the relation 
\begin{equation} 
-\frac{{\rm\bf K}^l{}_{l}}{\boldsymbol\kappa} \approx \frac{2(1+2H)}{1+H}
\end{equation}
holds on any of the $t=const$ Kerr-Schild time slices.

\section{Final remarks}\label{final}
\setcounter{equation}{0}

In this paper the constraints for smooth $[n+1]$-dimensional (with $n\geq 3$) Riemannian or Lorentzian spaces satisfying the Einstein equations were considered. The initial data surface was assumed to be foliated by a one-parameter family of codimension-one surfaces, which, for simplicity, were assumed to be orientable either compact without boundary in $\Sigma$ or non-compact and infinite. It was shown, regardless whether the primary space was Riemannian or Lorentzian, that the constraints can be put into the form of a coupled hyperbolic-parabolic system (if the mean curvature of the foliating surfaces is positive) or, alternatively, into a Friedrichs symmetrizable hyperbolic system and a subsidiary algebraic relation. In both cases those conditions which guarantee the (local) existence and uniqueness of smooth solutions to these evolutionary systems were identified. 

\medskip

Notably both of the above proposed evolutionary methods provide immediate answer to the first part of the questions raised in \cite{wald} (see also the second paragraph of the introduction of this paper) by providing a distinguished splitting of the basic variables $h_{ij}$ and $K_{ij}$, or that of $\hat N,\hat N^i,\hat \gamma_{ij}, \boldsymbol\kappa, {\rm\bf k}{}_{i}, {\rm\bf K}^l{}_{l},\interior{\rm\bf K}_{ij}$. For instance, in case of the hyperbolic-parabolic system $\hat N, {\rm\bf k}{}_{i}$ and ${\rm\bf K}^l{}_{l}$ are constrained whereas $\hat N^i,\hat \gamma_{ij},\boldsymbol\kappa$ and $\interior{\rm\bf K}_{ij}$ are freely specifiable throughout $\Sigma$. The initial data $({}_0\hat N, {}_0{\rm\bf k}_{i}, {}_0{\rm\bf K}^l{ }_{l})$, with ${}_0\hat N>0$, is also freely specifiable on one of the level surfaces of the foliation $\mycal{S}_{\rho}$ of $\Sigma$. Alternatively, in case of the symmetrizable hyperbolic system the variables $\boldsymbol\kappa, {\rm\bf k}{}_{i}$ and ${\rm\bf K}^l{}_{l}$ are 
constrained whereas $\hat N,\hat N^i,\hat \gamma_{
ij}$ and $\interior{\rm\bf K}_{ij}$ are freely specifiable throughout $\Sigma$. Note that the initial data for (\ref{constr_mom1_n}) and (\ref{ort_const_n}) is also at our disposal, i.e.~the values of ${}_0{\rm\bf k}{}_{i}$ and ${}_0{\rm\bf K}^l{}_{l}$ are freely specifiable on one of the level surfaces of the foliation $\mycal{S}_{\rho}$ of $\Sigma$. 

\medskip

Note that the method built on the first order symmetrizable hyperbolic system  (\ref{constr_mom1_n}) and (\ref{ort_const_n}) appears to be applicable in deducing initial data for black hole configurations \cite{racz_winicour}. One of the advantages of using the pertinent evolutionary setup is that---as opposed to the requirements of the conformal method where one needs to specify boundary data in the asymptotic region, as well as, in the strong field regime inside the black holes---it suffices to specify initial data only on a distant sphere surrounding the binary system in the asymptotic region. It is expected that the use of the proposed evolutionary approach will yield a reduction of the 
spurious gravitational wave content of numerical simulations.

\medskip

Note that as the momentum constraint takes the form of a first order symmetric hyperbolic system---see also analogous systems in \cite{racz_geom_det,racz_geom_cauchy}---the following interesting conceptual issue is raised. Namely, hyperbolicity in the conventional setup is always related to the causal structure determined by a, possibly evolving, Lorentzian spacetime metric. For the first glance the hyperbolicity of the momentum constraint, as given by (\ref{par_const_n})-(\ref{ort_const_n}), appears to be significantly different as---regardless of the signature of the primary space---the metric induced on $\Sigma$ is simply Riemannian that has nothing to do with causality. 

Notice, however, that by assuming the existence of a foliation $\mycal{S}_\rho$ of $\Sigma$ we also introduced a normal vector field $\hat n^i$ that is well-defined (apart from centers) on $\Sigma$. Taking then into account the well-known result that with the help of such a vector field it is always possible to define a Lorentzian metric $\check{h}_{ij}$ on $\Sigma$ via the relation  (see, e.g.~\cite{markus})
\begin{equation}\label{lor_riem}
\check{h}_{ij} = {h}_{ij} - (1+\alpha)\,\hat n_i \hat n_j\,,
\end{equation}
where $\alpha$ is a positive real function on $\Sigma$. (In particular, if $\alpha=1$ the vector field $\hat n^i$ gets to be a unit norm timelike vector field with respect to the Lorentzian metric $\check{h}_{ij}$.) 

Note that in virtue of (\ref{hij}) and (\ref{char_cone}), the latter determining the characteristic cone for the system (\ref{par_const_n})-(\ref{ort_const_n}), a Lorentzian metric $\check{h}_{ij}$, with $\alpha=n-1\geq 2$ (for $n\geq 3)$, entered implicitly into our discussions. This provides then an immediate explanation of the appearance of hyperbolicity in context of the momentum constraint.

\medskip

The linearity of the Hamiltonian constraint (\ref{scal_constr_n}) in $\boldsymbol\kappa$ may play important role in canonical quantization of gravity. For instance, in Dirac's approach that is known to apply to constrained systems such as Einstein's theory of gravity, the constraints are ignored at the classical level and the physical quantum states are supposed to be determined by solving the quantum constraints on a kinematical Hilbert space. It is of obvious interest to know whether the linearity of the Hamiltonian constraint in $\boldsymbol\kappa$ could simplify the identification of the true physical quantum states.  

\medskip

As already indicated in section \ref{constraints}, it would be of obvious interest to identify those conditions which could guarantee the global existence and uniqueness---and, possibly, the asymptotically Euclidean character and/or the regularity at centers---of solutions to the evolutionary systems introduced in this paper. 

\medskip

The clear up of all the above mentioned open problems, in particular, those in the last two paragraphs---and the involvement of foliations the level surfaces of which are compact with boundary in $\Sigma$ (see also footnote~\ref{FN})---would certainly deserve further attention.

\section*{Acknowledgments}

The author is grateful to Bob Wald and Jeff Winicour for helpful comments and suggestions. Thanks are due to the Albert Einstein Institute 
in Golm, Germany for its kind hospitality where parts of the reported results were derived. This research was also supported in parts by 
the Die Aktion \"Osterreich-Ungarn, Wissenschafts- und Erziehungskooperation grant 90\"ou1 and by the NKFIH grant K-115434.


\end{document}